\documentclass[12pt,preprint]{aastex}

\begin{document}

\title{The galactic disk thickness and the mass of the spherical
component}

\author{D.Bizyaev\altaffilmark{1,2}, A.Khoperskov\altaffilmark{3}, and N.Tiurina\altaffilmark{2}}

\altaffiltext{1}{ University of Texas at El Paso, US}
\altaffiltext{2}{Sternberg Astronomical Institute, Russia}
\altaffiltext{3}{ Dept. of Physics, Volgograd State University, Russia}

\begin{abstract}
We present results of numerical modeling made for 
the galactic stellar disk embedded in the spherical halo.  The 
non-linear dynamics of bending instabilities developed in the 
disk is studied.  The axisymmetrical  bending mode is 
considered as a main factor increasing the disk thickness. 
The model dependence between the disk thickness and the 
halo to disk mass ratio $M_h/M_d$ is inferred. The modeling 
results enable us to estimate the dark halo mass for several 
edge-on galaxies. The obtained values of $M_h/M_d$ for the 
galaxies considered are of order  1 - 2.
\end{abstract}

\noindent {\bf MODELING}

The dynamic modeling is based on numerical integration of the motion 
equations for N gravitationally interacting particles [1]. The system of 
collisionless particles forms the disk and bulge (if any) in its central     
part and is embedded into the dark halo. The large-scale bending 
instabilities in stellar disk are considered 
as one of the main factor regulating the disk thickness ([1],[2],[3]).

\noindent {\bf Model parameters:} 
\begin{itemize}
\item Spherical to disk mass ratio $M_s/M_d$
\item radial scale of the halo,  
\item the disk central surface density $\Sigma$.
\end{itemize}

\noindent {\bf Initial conditions} 

The initial disk is: 
\begin{itemize}
\item axisymmetrical, 
\item in an equilibrium along the radial and vertical directions, 
\item gravitationally stable in the disk plane
\end{itemize}

\noindent We COMPARE the output model parameters with the observed

\begin{itemize}
\item  rotation curve,
\item  the scale length $h$ and scale height $z$ of stellar disk
\item  the radial velocity dispersion (if available)
\end{itemize}
for edge-on galaxies (see Table 1). As a result, we obtaine the parameters
of the disk and dark halo (see Table 2). Two interesting relations between
the stellar disk thickness and defined parameters are shown in Fig 1 and 2.

\clearpage

\begin{table}
\begin{center}
\caption{Assumed parameters of the galaxies}
\begin{tabular}{lrrrrr}
\tableline\tableline
Name  &  D, Mpc & $h$, kpc & $z$, kpc \\
\tableline
UGC 6080 &  32.3 & 2.9 & 0.69  \\
NGC 4738 &  63.6 & 4.7 & 1.30  \\
UGC 9556 &  30.6 & 3.6 & 0.51  \\
UGC 9422 &  45.6 & 3.5 & 0.80  \\
NGC 5170 &  20.0 & 6.8 & 0.82  \\
UGC 8286 &   4.8 & 2.0 & 0.26  \\
UGC 7321 &  10.0 & 2.1 & 0.17  \\
NGC  891 &   9.5 & 4.9 & 0.99  \\
\tableline
\end{tabular}
\end{center}
\end{table}

\begin{table}
\begin{center}
\caption{Parameters obtained with the help of N-body modeling}
\begin{tabular}{lrr}
\tableline\tableline
Name  &  $\Sigma$, $M_{\odot}/pc^2$ & $M_s/M_d$\\
\tableline
UGC 6080 &    700 & 1.0 \\
NGC 4738 &    970 & 0.5 \\
UGC 9556 &    510 & 1.1 \\
UGC 9422 &   1000 & 0.8 \\
NGC 5170 &    460 & 2.2 \\
UGC 8286 &    200 & 1.6 \\
UGC 7321 &    310 & 2.2 \\
NGC  891 &    530 & 1.0 \\
\tableline       
\end{tabular}    
\end{center}     
\end{table}     

\clearpage

\section*{CONCLUSIONS}

\noindent The parameters of stellar disks and dark halo are found for the
sample of eight edge-on spiral galaxies with the help of numerical N-body
modeling. The bending instabilities are considered as a main factor
increasing the disk thickness.
\medskip

\noindent The relation between the stellar disk thickness and the relative
mass of dark halo is shown for the galaxies studied. Since all but the one
galaxies are bulgeless, the spherical component mass reveals the dark halo
mass in our galaxies. The relation between the disk central surface density
and its relative thickness $z/h$ is presented.

\begin{figure}
\plotone{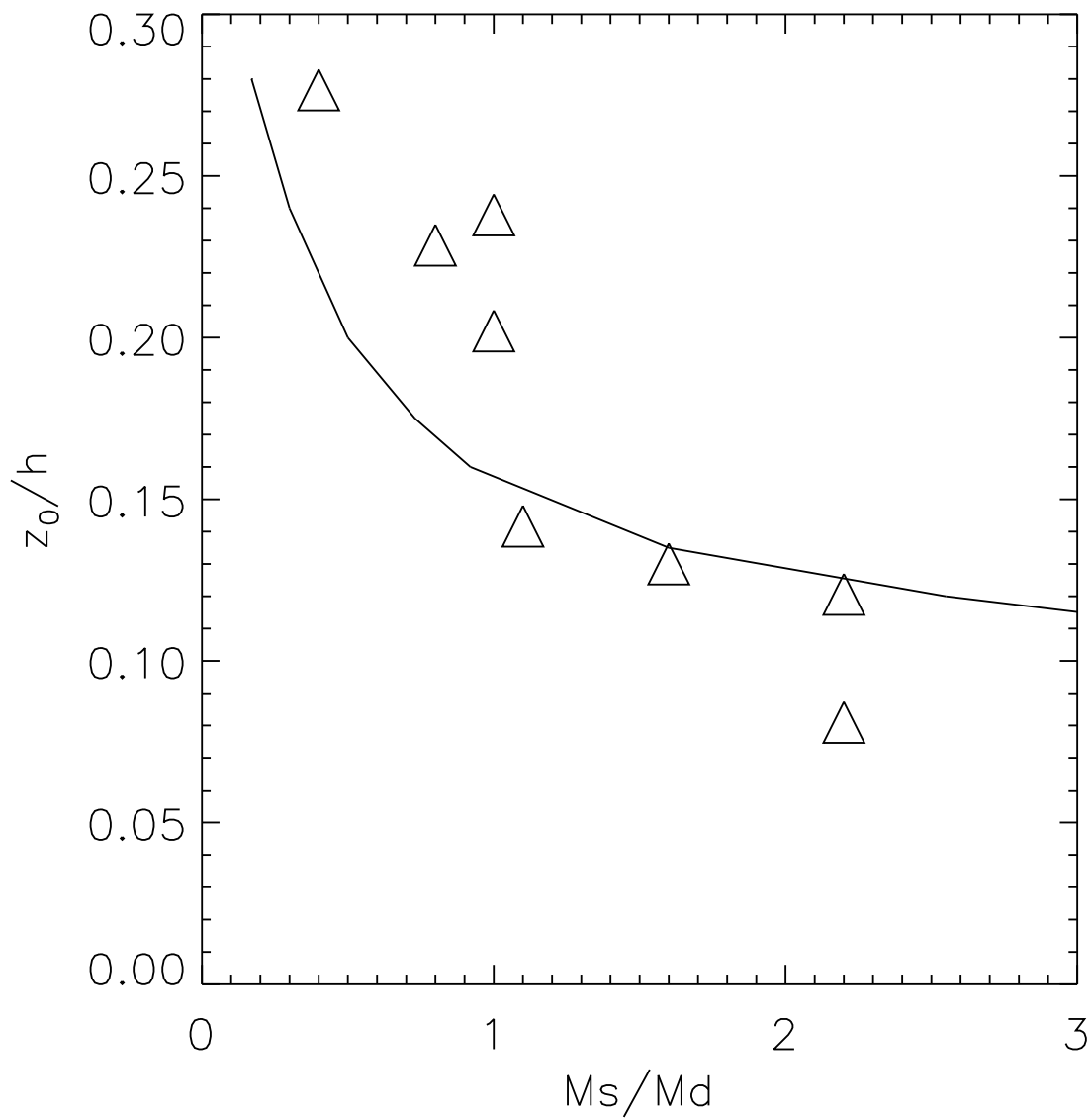}
\caption{The dependence between the spherical to disk mass ratio $M_s/M_d$ and the
relative stellar disk thickness $z/h$ is shown for our sample of galaxies by
triangles. The solid line indicates the same dependence made with the help
of N-body modeling (taken from [1]). Since all (except two) galaxies studied
are bulgeless, $M_s/M_d$ reveals the dark halo to disk mass ratio $M_h/M_d$.
\label{fig1}}
\end{figure}

\begin{figure}
\plotone{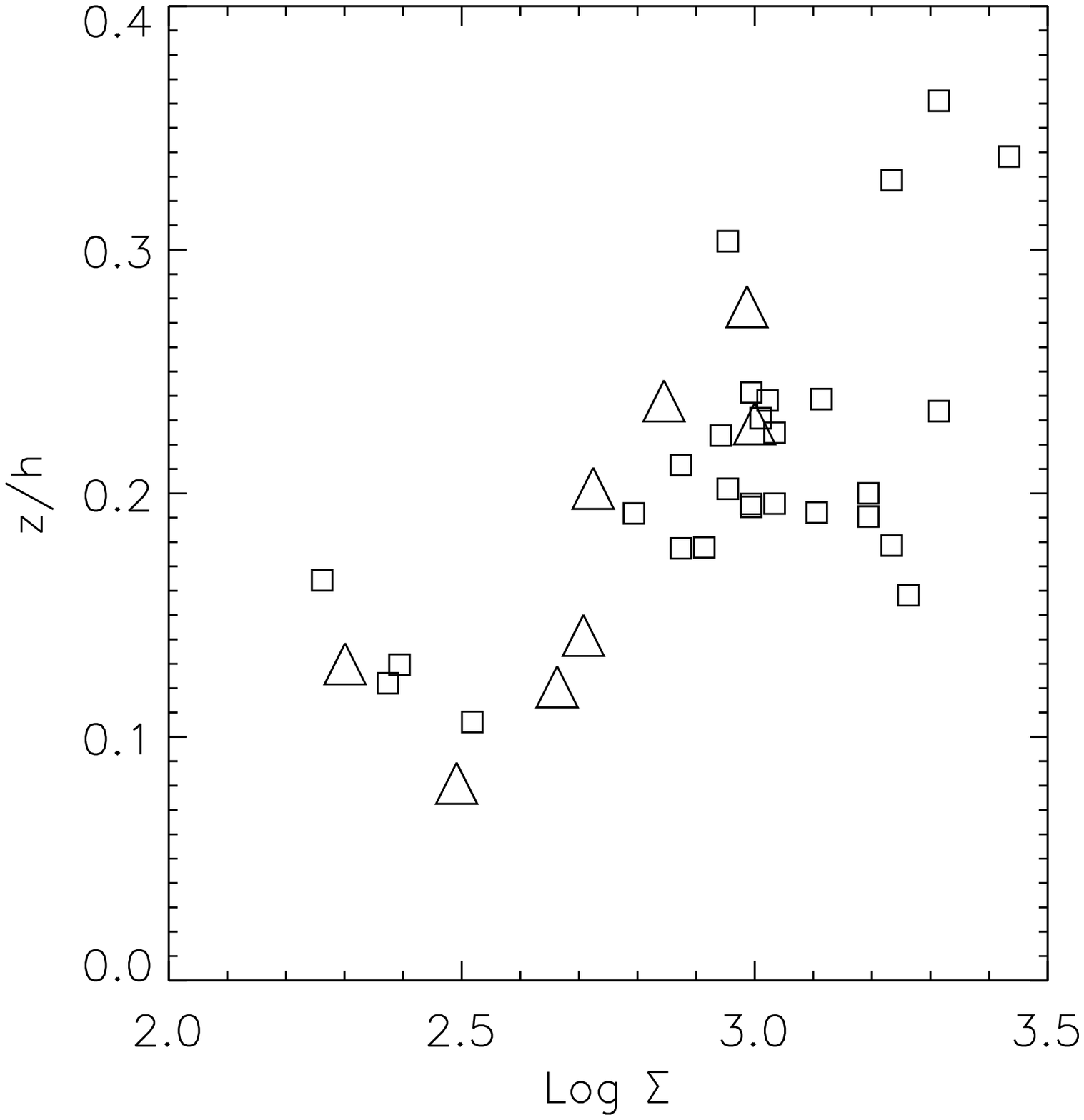}
\caption{The relation between the central surface density of stellar disk 
$\Sigma$ and its relative thickness $z/h$ is shown by the triangles for our 
galaxies.  The squares show the same parameters taken for the galaxies from 
[5].  The values of $\Sigma$ for those galaxies is obtained using a uniform 
mass to light value to coinside with the triangles.
\label{fig2}}                                                     
\end{figure}

\begin{acknowledgments}

DB is supported by the NASA/JPL grant NRA 99-04-OSS-058.
The project is partially supported by RFBR through the grant 01-02-17597

\end{acknowledgments}

REFERENCES

\noindent [1] Mikhailova E., Khoperskov A., Sharpak S., 2000, 
in conf. proc. ""Stellar Dynamics: from classic to modern"", 
ed. by Ossipkov \& Nikiforov, p.147 

\noindent [2] Hunter C., Toomre A., 1969, ApJ, 155, 747 

\noindent [3] Polyachenko V., Shukhman I., 1979, AZh, 56, 724

\noindent [4] Vandervoort P., 1991, ApJ, 377, 49

\noindent [5] Bizyaev D, Kajsin S., 2003, AAS 201 146.05, astro-ph/0306190

\end{document}